\documentclass{edbk}
\usepackage{epsf}

\normallatexbib

\begin{document}

\articletitle[Two-proton correlations at SPS]
{Two-proton correlations\\
from Pb+Pb central collisions}
 
\author{F.~Wang (for the NA49 Collaboration)}

\affil{Nuclear Science Division, Lawrence Berkeley National Laboratory\\
One Cyclotron Road, Berkeley, CA 94720, USA}
\email{FQWang@lbl.gov}

\begin{abstract}
The two-proton correlation function at midrapidity from Pb+Pb central
collisions at 158 AGeV has been measured by the NA49 experiment.
The preliminary results are compared to model predictions from proton 
source distributions of static thermal Gaussian sources and
the transport models of {\sc rqmd} and {\sc venus}.
We obtain an effective proton source size 
$\sigma_{\rm eff}=4.0\pm 0.15{\rm (stat.)}^{+0.06}_{-0.18}{\rm (syst.)}$~fm.
The {\sc rqmd} model underpredicts the correlation function
($\sigma_{\rm eff}=4.41$~fm),
while the {\sc venus} model overpredicts the correlation function
($\sigma_{\rm eff}=3.55$~fm).
\end{abstract}


\vspace{-2.6in}
\epsfysize=0.5in\epsfbox[25 0 145 40]{fqwang_null.ps}
\vspace{2.2in}

Nuclear matter under extreme conditions of high energy density
has been extensively studied through high energy heavy ion collisions.
The baryon density plays an important role in the 
dynamical evolution of these collisions.
To measure the baryon spatial density, one needs information on the 
space-time extent of the baryon source.
The space-time extent of the proton source at freeze-out can be inferred 
from two-proton correlation functions. The peak in the correlation
function at $q_{\rm inv}=\sqrt{-q_{\mu}q^{\mu}}/2 \approx 20$~MeV/$c$
is inversely related to the effective size of the proton 
source~\cite{Koo77:plb_pp,Gel90:rmp_corr}.
In the above, $q_{\mu}$ is the difference of the proton 4-momenta,
and $q_{\rm inv}$ is the momentum magnitude of one proton in the rest frame
of the pair.

We report the first, preliminary results on the two-proton correlation
function in the midrapidity region from Pb+Pb central collisions at 158 GeV 
per nucleon. The measurement was done by the NA49 experiment~\cite{na49_nim}
at the SPS, using the 5\% most central events,
corresponding to collisions with impact parameter $b\leq 3.3$~fm.

Two independent analyses were performed on the data:
$dE/dx$ analysis which used particles in the rapidity range $2.9<y<3.4$
(assuming proton mass) with at least 70\% probability to be a proton
obtained from their ionization energies deposited in the time projection
chambers (TPCs), and TOF analysis which used identified protons by
combining the time of flight information and the $dE/dx$.
Both analyses used protons up to a transverse momentum $p_T=2$ GeV/$c$.

The proton samples are contaminated by weak decay protons
($\Lambda+\Sigma^0$ and $\Sigma^+$) which are incorrectly
reconstructed as primary vertex tracks. From the measured single
particle distributions~\cite{App98:na49_prl_baryon,Bor97:sqm97} and
model calculations of {\sc rqmd} and {\sc venus}, we estimate 
the contamination to be $15^{+15}_{-5}\%$.
This results in about $30^{+20}_{-10}\%$ of the proton pairs having 
at least one proton from weak decays.
These protons do not have correlation with protons from the primary
interactions.
In the $dE/dx$ analysis, further contamination is present from kaons 
on the lower tail of their $dE/dx$ distribution merging into the region 
where particles have at least 70\% probability to a proton.
This resulted in 25\% $K^+p$ pairs and fewer than 2\% $K^+K^+$ pairs 
in the proton pair sample. The coulomb hole of $K^+p$ pairs
does not affect the resulting two-proton correlation function in the
interested low $q_{\rm inv}$ region below 50~MeV/$c$.

The two-proton correlation function is obtained as the ratio of 
the $q_{\rm inv}$ distribution of true proton pairs to that
of mixed-event pairs with protons from different events.
The number of mixed-event pairs was large enough so that
the statistical error on the correlation function is dominated
by the statistical uncertainty in the number of true pairs.
To eliminate the effect of close pair reconstruction inefficiency,
a cut of 2~cm was applied on the pair distance at the middle plane of
the TPC for both true and mixed-event pairs.

\begin{figure}[hbt]
\epsfysize=4in\epsfbox[80 140 500 640]{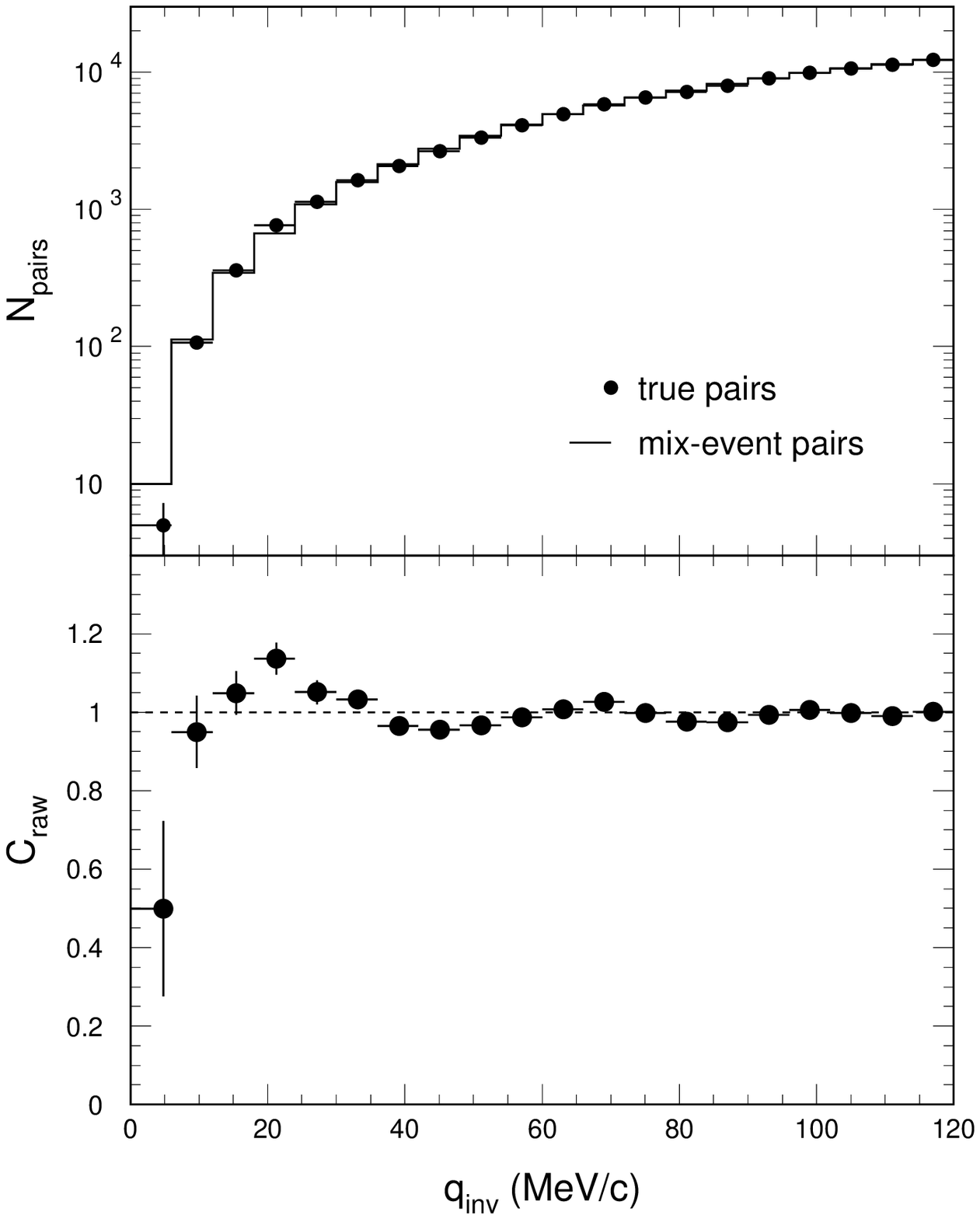}
\vspace{-2.5in}
\narrowcaption{Top: $q_{\rm inv}$ distributions of true proton pairs (points) 
	and mixed-event proton pairs (histogram).
	Bottom: the measured two-proton correlation function.
	The contamination from weak decay protons and the finite momentum 
	resolution were not corrected.
	The errors shown are statistical only.\label{fig:data}}
\vspace{0.3in}
\end{figure}

The correlation functions obtained from the two analyses can be directly 
compared because of the nearly symmetric acceptances about midrapidity
$y_{\rm c.m.}=2.9$.
The correlation functions (with the $K^+p$ contamination corrected in
the $dE/dx$ analysis) are consistent.
In the results reported below, the true pairs and the respective
mixed-event pairs from the two analyses were combined.
The combined sample had about $10^5$ pairs with $q_{\rm inv}<120$~MeV/$c$,
75\% of which were from the TOF analysis.
The $q_{\rm inv}$ distributions of the true pairs and the mixed-event pairs
are shown in the top panel of Fig.~\ref{fig:data}.
The number of mixed-event pairs was normalized to that of true pairs 
in the range $q_{\rm inv}>500$~MeV/$c$.
The resulting correlation function
is shown in the bottom panel of Fig.~\ref{fig:data}.
The prominent peak at $q_{\rm inv}\approx 20$~MeV/$c$ is evident.
There is a statistically significant structure in the correlation function 
at $q_{\rm inv}\approx 70$~MeV/$c$. 
Many systematic effects have been studied; 
none has been identified that can account for the structure.
There have been suggestions that a sharp edge in the 
two-proton density distribution of the source can produce such an
effect~\cite{sharp_edge}.

\begin{figure}[hbt]
\epsfysize=2.8in\epsfbox[50 180 620 590]{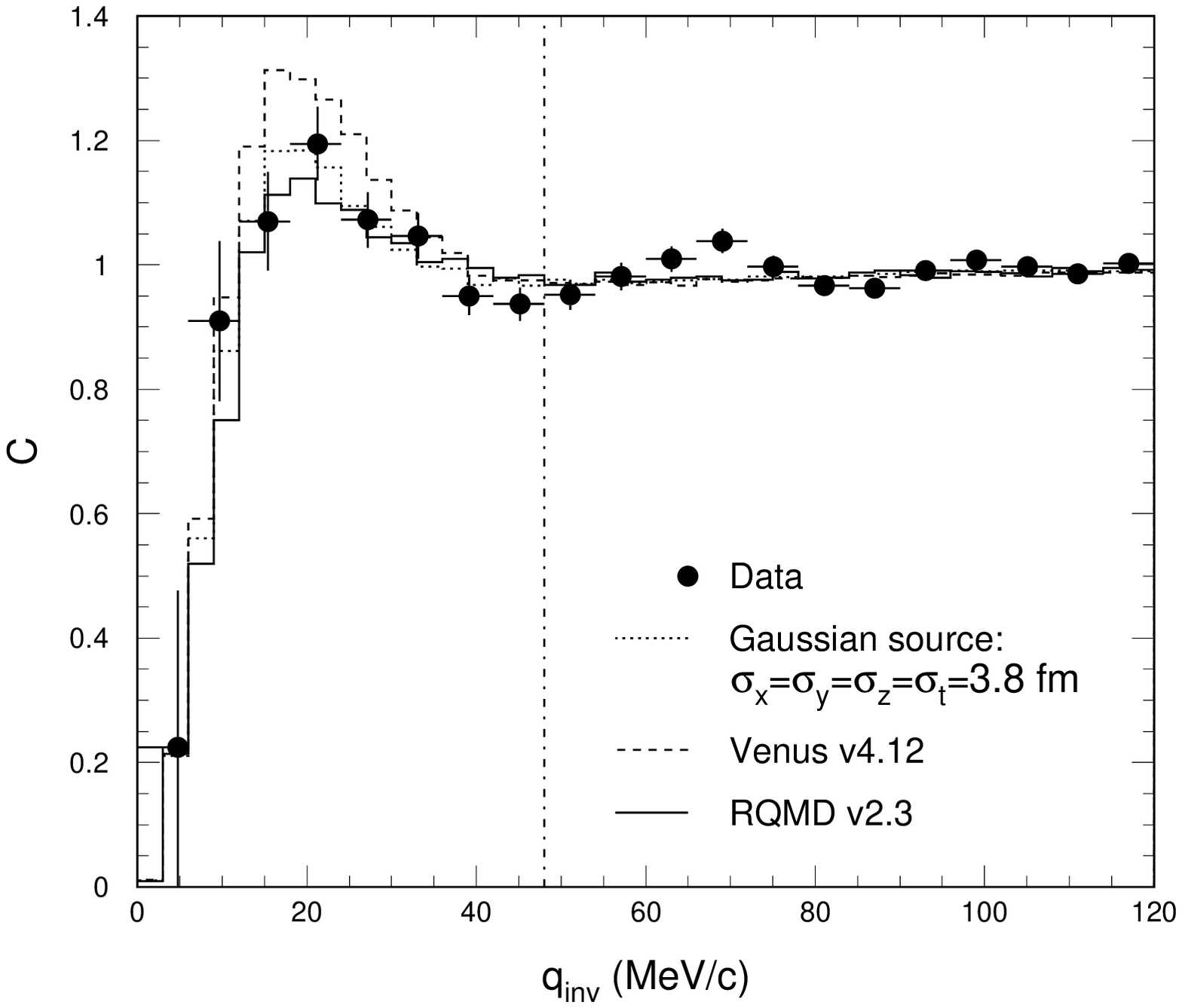}
\vspace{-3in}
\narrowcaption{The two-proton correlation function after corrections for
	the 30\% contamination due to weak decay protons and the finite 
	momentum resolution (points), compared to calculations for
	a Gaussian source (dotted), and for freeze-out protons from 
	{\sc rqmd} v2.3 (solid) and {\sc venus} v4.12 (dashed). The errors
	shown on the data points are statistical only.\label{fig:model}}
\vspace{0.2in}
\end{figure}

We correct the measured correlation function for the contamination 
from weak decay protons and for the finite momentum resolution.
The effect of the momentum resolution is only significant in the first
two data points of the measured correlation function. 
The corrected correlation function is plotted 
in Fig.~\ref{fig:model} as filled points.

In order to assess the proton freeze-out conditions, we compare the 
measured two-proton correlation function to theoretical calculations.
Given the proton phase space density distribution, the two-proton 
correlation function can be calculated by the Koonin-Pratt 
Formalism~\cite{Koo77:plb_pp,Pra87:two_proton}.
Two types of proton source were used:

(I) Isotropic Gaussian sources of 
widths $\sigma_{x,y}, \sigma_z$ and $\sigma_t$
for the space and time coordinates of protons in the source rest frame,
and thermal momentum distribution of temperature $T$.
No correlation between space-time and momentum of the protons is present.
Following combinations of parameters were used in the calculations:
 $\sigma_{x,y}=\sigma_z=\sigma$, $\sigma_t=0$ and $\sigma$,
and $T=120$~MeV (as derived in~\cite{App98:epj_expansion}), 
300~MeV (measured inverse slope of proton transverse mass 
spectrum~\cite{App98:na49_prl_baryon})
and 70~MeV (inverse slope observed at low energy, as an extreme).

(II) Protons generated for Pb+Pb central collisions ($b\leq 3.3$~fm)
at 158 AGeV by two microscopic transport models:
the Relativistic Quantum Molecular Dynamics ({\sc rqmd}) model 
(version 2.3)~\cite{rqmd}
and the {\sc venus} model (version 4.12)~\cite{Wer93:venus}.
Both models describe a variety of experimental data on single particle
distributions reasonably well.
Protons at freeze-out have correlations between space-time
and momentum intrinsic to the dynamical evolution in the models.

Only protons in the experimental acceptance are used
to calculate the two-proton correlation functions,
the results of which are shown in Fig.~\ref{fig:model} for
{\sc rqmd}, {\sc venus}, and the Gaussian source 
with $\sigma_{x,y}=\sigma_z=\sigma_t=3.8$~fm and $T=120$~MeV, respectively.
While {\sc venus} overpredicts the amplitude of the 
correlation function, {\sc rqmd} slightly underpredicts the amplitude.

We use $\chi^2$/ndf, the normalized mean square of the point-to-point
difference between the data and the calculation in the 
range $q_{\rm inv}<48$~MeV/$c$ ({\it i.e.}, 8 data points or ${\rm ndf}=8$),
to quantify how well the calculations agree with data.
We characterize the effective size of the proton source from each model
by $\sigma_{\rm eff} = 
 \sqrt[3]{\sigma_{\Delta x}\cdot\sigma_{\Delta y}\cdot\sigma_{\Delta z}}/
 \sqrt{2}$,
where $\sigma_{\Delta x}, \sigma_{\Delta y}$ and $\sigma_{\Delta z}$ 
are the Gaussian widths fitted to the distributions in $\Delta x, \Delta y$
and $\Delta z$, distance between the protons of 
close pairs with $q_{\rm inv}<48$~MeV/$c$.
The distance is evaluated in the pair rest frame at the time 
when the later particle freezes out~\cite{note_2m}.
Respectively for {\sc rqmd} and {\sc venus},
the $\chi^2$/ndf values are 1.47 and 2.29; the fitted Gaussian widths are
($\sigma_{\Delta x},\sigma_{\Delta y},\sigma_{\Delta z})=(5.91,6.00,6.83)$~fm
 and $(4.57,4.57,6.08)$~fm, where $z$ is the longitudinal coordinate;
the resulting effective sizes are $\sigma_{\rm eff}=4.41$~fm and $3.55$~fm.

\begin{figure}[hbt]
\epsfysize=2.8in\epsfbox[50 180 620 590]{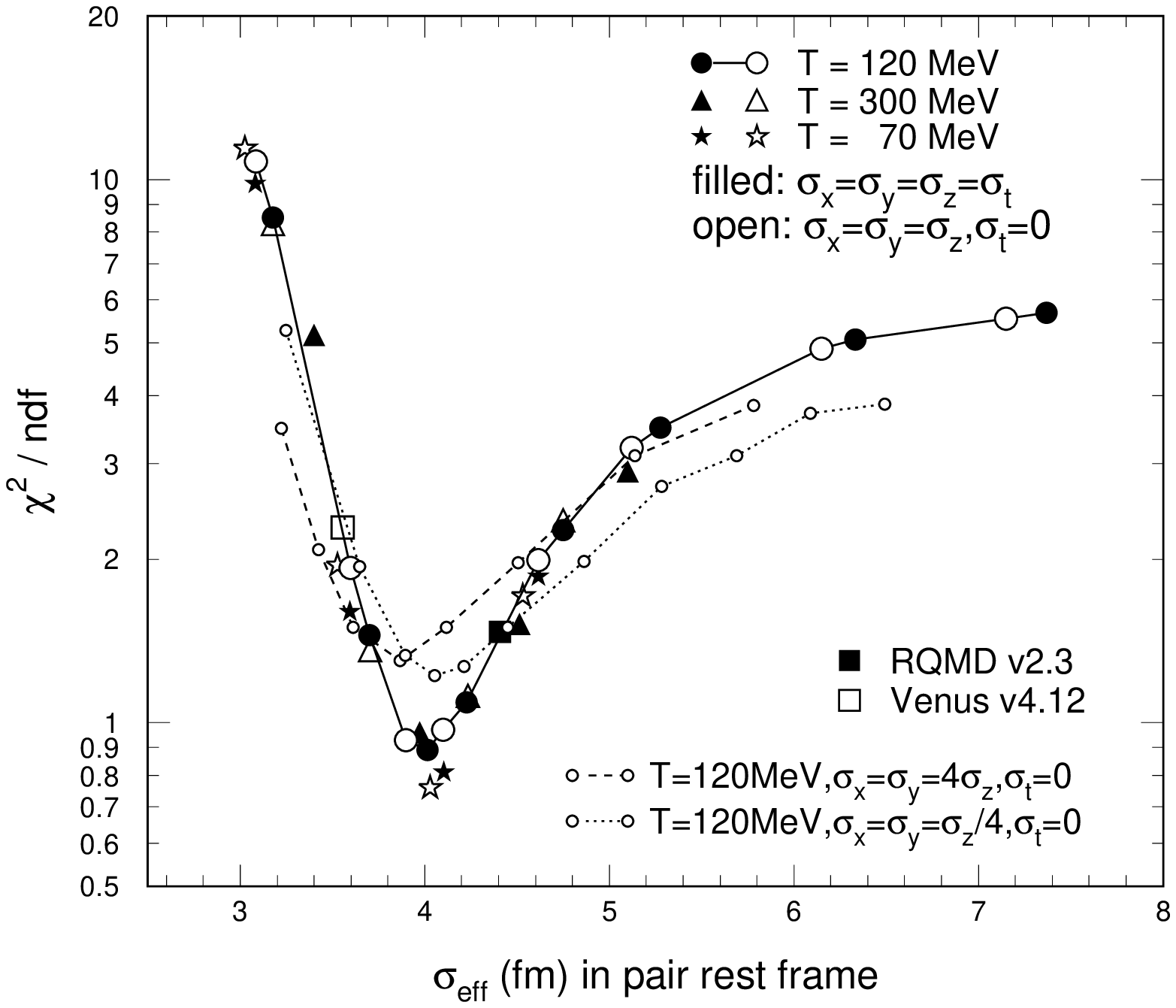}
\vspace{-2.85in}
\narrowcaption{The $\chi^2$/ndf values as function of the effective source
	size $\sigma_{\rm eff}$ for various model calculations with respect 
	to the measured correlation function. The model calculations
        are for Gaussian sources (circles, triangles, and stars) and 
	for freeze-out protons generated by {\sc rqmd} v2.3 (filled square)
	and {\sc venus} v4.12 (open square).\label{fig:chi2}} 
\vspace{0.2in}
\end{figure}

In Fig.~\ref{fig:chi2}, we study the $\chi^2$/ndf as a function
of $\sigma_{\rm eff}$.
The $\chi^2$/ndf values from all three models follow roughly the same 
solid line, drawn through the points for the Gaussian sources
with $T=120$~MeV to guide the eye~\cite{note_curve}.
From the minimum $\chi^2$/ndf point and the points where $\chi^2$/ndf has
increased by 0.125 (note ${\rm ndf}=8$), 
we extract $\sigma_{\rm eff}=(4.0\pm 0.15)$~fm,
where 0.15~fm is the statistical error~\cite{PDG98:error}.
By applying a correction to the measured correlation function using a proton 
pair contamination of 20\% and 50\%, we obtain a systematic error 
of $^{+0.06}_{-0.18}$~fm on $\sigma_{\rm eff}$.
We note that $\sigma_{\rm eff}=4.0$~fm corresponds to a uniform density
hard sphere of radius $\sqrt{5}\sigma_{\rm eff}=8.9$~fm.

We have also studied Gaussian sources with extreme shapes: 
oblate $\sigma_{x,y}=4\sigma_z$ and 
prolate $\sigma_{x,y}=\sigma_z/4$ (both with $\sigma_t=0$).
The corresponding $\chi^2$/ndf versus $\sigma_{\rm eff}$ curves
are shown in Fig.~\ref{fig:chi2} as dashed and dotted lines, respectively. 
They do not fall along the solid line, implying
that the two-proton correlation function has certain 
sensitivity to the shape of the proton source.
In principle, multi-dimensional two-proton correlation functions
could reveal the shape of the proton source. However,
the lack of statistics has prevented us from drawing conclusions.

We note that there is no simple relation between $\sigma_{\rm eff}$
and the proton source in {\sc rqmd} and {\sc venus} model.
From single proton distributions at freeze-out, we obtain the following 
Gaussian widths in the source rest frame:
($\sigma_x,\sigma_y,\sigma_z,\sigma_t)=(7.6,7.7,6.4,7.0)$~fm
for {\sc rqmd} and $(3.6,3.6,4.3,1.9)$~fm for {\sc venus}.
The two-proton correlation function, therefore, appears to measure a smaller 
region of the source, {\it i.e.}, $\sigma_{\Delta i}<\sqrt{2}\sigma_i$ 
($i=x,y,z$), due to space-time-momentum correlation.
The effect is more dramatic in {\sc rqmd} than in {\sc venus}, 
which is consistent with the expectation that more secondary 
particle interactions in {\sc rqmd} result in a stronger correlation 
between space-time and momentum of freeze-out protons.
The fact that the $\chi^2$/ndf values versus $\sigma_{\rm eff}$ for 
{\sc rqmd} and {\sc venus} lie on the curves obtained for the 
isotropic Gaussian sources, in which no space-time-momentum correlation
is present, suggests that the effect of the space-time-momentum correlation
is small in $\sigma_{\rm eff}$. 

Finally, we comment on our two-proton correlation function
in the context of other measurements.
The pion source size measured by interferometry increases with
the pion multiplicity~\cite{hbt_ref},
which increases steadily with bombarding energy in similar colliding
systems~\cite{Gaz95:pion}.
Due to the large pion-nucleon cross-section, one would expect that protons
and pions freeze-out under similar conditions, therefore, the proton source 
size would increase with bombarding energy as well.
However, our measurement, in conjunction with preliminary results obtained
at GSI~\cite{gsi} and AGS~\cite{e895_prelim} energies,
shows that the peak height is rather insensitive to the bombarding energy.
This implies that the effective sizes of the freeze-out proton sources 
are similar in heavy ion collisions over a wide energy range.
More detailed studies are needed to understand the possible acceptance
and instrumental effects in these measurements.

In summary, the NA49 experiment has measured the two-proton
correlation function at midrapidity from Pb+Pb central collisions
at 158 GeV per nucleon.
From comparisons between the data and the calculations,
we extract an effective proton source size of
$\sigma_{\rm eff}=4.0\pm 0.15{\rm (stat.)}^{+0.06}_{-0.18}{\rm (syst.)}$~fm.
The {\sc rqmd} model underpredicts the amplitude of the correlation function
($\sigma_{\rm eff}=4.41$~fm),
while the {\sc venus} model overpredicts the amplitude
($\sigma_{\rm eff}=3.55$~fm).
Due to the space-time-momentum correlation, the two-proton correlation 
function is sensitive only to a limited region of the proton source.
Our measurement together with the measurements at lower energies suggest
a very weak dependence of the two-proton correlation function on
bombarding energy. 

Acknowledgment: I would like to thank M.~Cristinziani who did part of the
data analysis. Fruitful discussions with Drs. P.M.~Jacobs, R.~Lednicky, 
S.~Panitkin, A.M.~Poskanzer, H.G.~Ritter, P.~Seyboth, S.~Voloshin and
N.~Xu are greatly acknowledged.
This work was partially supported by the Director, Office of Energy Research, 
Division of Nuclear Physics of the Office of High Energy and Nuclear Physics 
of the US Department of Energy under Contract DE-AC03-76SF00098.


\begin{chapthebibliography}{10}
\bibitem{note_2m}
For a chaotic Gaussian source, $\sigma_{\rm eff}$ is close to the
  non-relativistic approximation of the effective size in the source rest
  frame, $\sqrt[6]{(\sigma_{x,y}^2+\frac{T}{2m}\sigma_t^2)^2\cdot
  (\sigma_z^2+\frac{T}{2m}\sigma_t^2)}$, where $\frac{T}{m}$ is the
  one-dimensional mean squared velocity of the thermal protons, and
  $\frac{T}{2m}$ is that of the close pairs.
\bibitem{note_curve}
The shape of the $\chi^2/{\rm ndf}$ versus $\sigma_{\rm eff}$ curve can be
  understood as follows: the $\chi^2$/ndf approaches an asymptotic value
  for large sources to which the two-proton correlation function is not 
  sensitive to the source size any more, whereas for small sources, the
  two-proton correlation function is very sensitive and its strength 
  decreases rapidly with increasing source size.
\end{chapthebibliography}
\end{document}